\documentclass{optica-article}

\journal{opticajournal} 

\articletype{Research Article}

\begin{document}

\title{Coherent manipulation of the biphoton generation in cavity-QED system}

\author{Jia-Ni Yang\authormark{},
Xin-Yi Ling\authormark{},
Yuan Feng\authormark{},
Xiao-Jun Zhang\authormark{*},
and Jin-Hui Wu\authormark{$\dagger$}}

\address{\authormark{}School of Physics and Center for Quantum Sciences, Northeast Normal University, Changchun 130024, China}

\email{\authormark{*}zhangxj037@nenu.edu.cn} 
\email{\authormark{$\dagger$}jhwu@nenu.edu.cn} 


\begin{abstract*} 
We theoretically investigate the coherent manipulation of biphoton generation via spontaneous four-wave mixing in a cavity-QED system with a single atom. The atom is driven by pumping, coupling, and driving fields, and the generation of the Stokes and anti-Stokes photons are enhanced by two cavities. By solving the master equation in the steady state, we analyze the spectral brightness, as well as the degree of the auto-correlation and cross-correlation. Our results show that when the pumping and driving fields are in two-photon resonance, the dark state established between the ground and Rydberg states. efficiently enhances the controllability of the driving field over the biphoton generation and the quantum statistics. 
In contrast, under large two-photon detuning, the control capability of the driving field is significantly reduced. The coupling field, which directly relates to the electromagnetically induced transparency, modifies the linewidth of the biphoton, while the atom-cavity coupling strength only changes the brightness without affecting the linewidth.

\end{abstract*}

\section{Introduction}
 In the quantum domain beyond classical boundaries, nonclassical photon pairs serve as standard tools for both exploration and application. For a quantum network whose atomic nodes are connected via flying single photons, the network's ability to achieve quantum connectivity and scalability hinges directly on the strength of atom-photon interactions \cite{duan2001long, kimble2008quantum, ritter2012elementary}, that includes the efficiency of the biphoton generation from the nonclassical light source. 

Spontaneous four-wave mixing (SFWM) based on electromagnetically induced transparency (EIT) \cite{RevModPhys.77.633}, which produces biphoton states with a narrower bandwidth \cite{PhysRevLett.94.183601, Zhao:14,WOS:001389473100002,PhysRevLett.97.113602}, draws considerable attention today \cite{PhysRevResearch.4.023132,Shih:24,p84v-1xqv}. In a standard scheme, this kind of SFWM, also the only kind we focus on in this paper, is realized in atomic medium driven by the pumping and coupling fields, and the third-order nonlinear process is spontaneously initiated, leading to the generation of the biphotons. 
The two photons are normally call Stokes photon and anti-Stokes photon. The anti-Stokes photon survives absorption with the help of the EIT effect induced by the coupling field, while the Stokes photon is preserved due to the large detuning of the pumping field.
The two photons are endowed with strong spectral and temporal correlation, and this feature is of great importance to the field of quantum cryptography \cite{p84v-1xqv,WOS:000385429400004,WOS:000640217000011}, quantum imaging \cite{WOS:000340840600028,WOS:001187580500004,WOS:001048092600002,WOS:000614789700002}, quantum metrology \cite{PhysRevLett.132.180802,WOS:000983415900015,PhysRevLett.130.070801} and quantum computation \cite{WOS:000970707400001,WOS:001216366800026,WOS:000166175600033}.

The generation of the biphotons from the atomic ensembles can be viewed as the act of macroscopic polarizability, with the spontaneous process triggered by the free-space vacuum. To enhance the interaction between the atoms and the field, one could place the atomic medium in cavity \cite{PhysRevLett.101.190501,10.1063/5.0006021, PhysRevA.111.053705}. In such systems, photon pairs are triggered by the cavity mode, and with the enhancement from the reduction of the mode volume, a single atom could be enough for considerable generation rate.



\begin{figure}[htbp]
\centering
\includegraphics[width=0.65\linewidth]{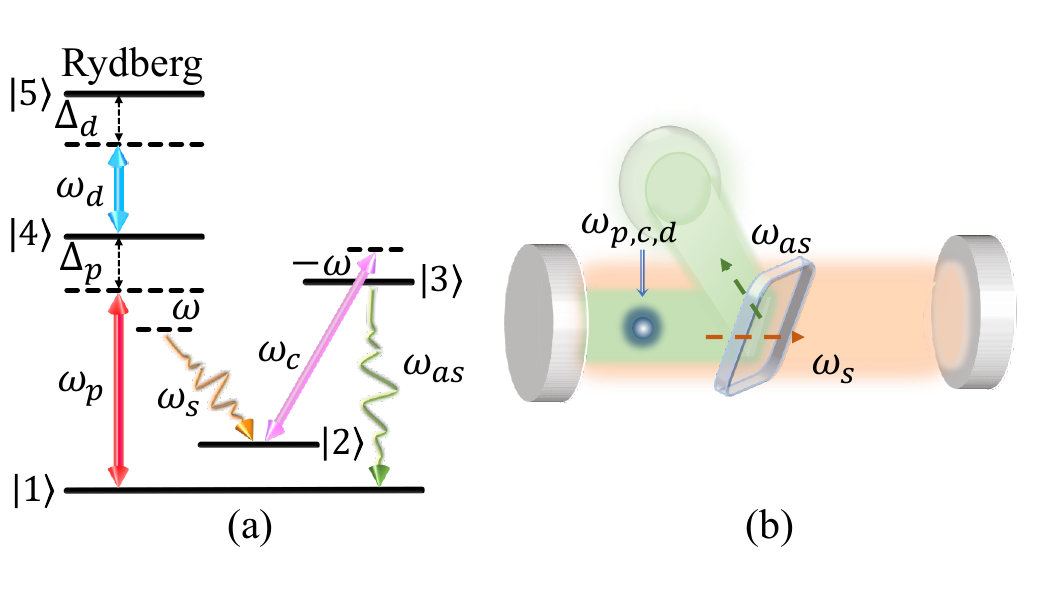}
\caption{\label{fig1} (a) Five-level $^{87}$Rb atom driving by pumping field $(\omega_{p})$, coupling field $(\omega_{c})$ and driving field $(\omega_{d})$ to generate the photon pair $(\omega_{s},\omega_{as})$. (b) Schematic for spontaneous biphoton generation controlled by the three fields.}
\end{figure}

In general, the biphoton generation rate and statistics can be coherently controlled by applying an additional laser, with a properly designed configuration of interaction. Such controlling  enables on-demand, programmable quantum light sources, and could lead to a vast range of applications, for example, in integrated optics 
\cite{PhysRevLett.130.223601,He:24} and all-optical quantum logic \cite{WOS:000393737500035, WOS:000537039500005}. In designing, one should always avoid unwanted nonlinear interactions introduced by this newly added laser field, which may degrade the quality of the biphoton.

In this paper, we investigate a single-atom SFWM system in cavities. The highest excited state of the atom is a Rydberg state. Rydberg levels refer to the highly excited electronic energy levels of atoms (or molecules) where a valence electron occupies an orbital with an extremely large principal quantum number. Rydberg levels exhibit extraordinary properties, such as large electric dipole moments, small energy level spacings and the long radiative lifetimes. These characteristics endow Rydberg atoms with significant application prospects in research areas such as quantum information processing\cite{shi2025barium,palaiodimopoulos2024chiral}, quantum computing\cite{begoc2025controlled,Hou:25}, and quantum precision metrology\cite{liang2026exceptional}. Our model uses the feature of the long radiative lifetimes, and it helps us to introduce atomic coherence under the application of the (additional) driving field. It can manipulate the biphoton generation effectively, but with minimal additional decay processes.

We use master equation to describe the dynamical evolution, and solve it for the steady-state spectral brightness and correlations for the Stokes and anti-Stokes photon. The results show that the dark state between the ground and Rydberg states dramatically influence the control capability of the driving field. When the pumping and driving field are double-photon resonant with the transition $|1\rangle \leftrightarrow |5\rangle$ and a dark state is built, the generation of the biphotons and their statistics appear quite sensitive to the strength of the driving field. On the other hand, if the pumping and driving fields are double-photon large detuned, the control capacity of the driving field is effectively reduced. We also investigate the influence of the coupling field and the coupling coefficient between the atom and cavities (equivalent to cavity volume). We find that the coupling field, directly related to the EIT effect, can change the linewidth of the biphoton with fixed decay rate from the cavity, while the cavity volume only changes the brightness, and leaves the linewidth as a constant.

The structure of this paper is organized as follows. In Sec. II, the dynamics of the atomic system interacting with applied fields and the environment are described by solving the corresponding master equation. In Sec. III we discuss the role of the relevant subsystems and the dark state that could be built by the pumping and driving fields. In Sec. IV, we show how the biphoton generation changes according to the driven field, coupling field and the coefficient of interaction between the system and environment. We summarize the main conclusions of this paper in Sec. V. 

\section{Models and equations}

Let us consider a five-level atom, as shown in Fig.\ref{fig1}(a) driven by three applied fields at frequencies $\omega_p$, $\omega_c$ and $\omega_d$, respectively. The pumping field ($\omega_p$) and the coupling field ($\omega_c$) are the necessary energy source for the SFWM process that generates the Stokes photon at $\omega_s$ and simultaneously the anti-Stokes photon at $\omega_{as}$. As a parametric nonlinear process, SFWM would always generate the Stokes and anti-Stokes photons in pairs. 
The driving field ($\omega_d$) connects the atomic level $|4\rangle$ and the Rydberg state $|5\rangle$, whose frequency and amplitude could serve as the ``control knob'' to tune the generation of the photon pairs. The long lifetime of the Rydberg state ensures that the spontaneous decay from it does not hamper the correlation between the Stokes and anti-Stokes photon. In addition, the strong coherence could be built between $|1\rangle$ and $|5\rangle$ which brings the system into the dark state. In the following analysis, we demonstrate that the dark state is a necessary condition for the effective control from the driving field.

The atom is assumed to be seated in a double-cavity system as shown in Fig. \ref{fig1} (b). The Stokes photon is generated as a quanta of a particular mode of the Stokes cavity, while the anti-Stokes as the other. For simplicity, we assume that the photon resonates with its corresponding cavity. All the other applied fields travel along the perpendicular direction. Since we only have one atom for the nonlinear energy conversion, the phase-matching condition is not necessarily required. The Hamiltonian of the interaction in interaction picture can be written as
\begin{equation}
\begin{split}
\frac{\hat{V}}{\hbar}=&-\Delta_{p}\hat{\sigma}_{44}- \delta \hat{\sigma}_{55}-\omega(\hat{\sigma}_{33}+\hat{\sigma}_{22})\\
&- (g_{s}\hat{a}_{s}\hat{\sigma}_{42} +g_{as}\hat{a}_{as}\hat{\sigma}_{31}+\Omega_{p}\hat{\sigma}_{41}+\Omega_{c}\hat{\sigma}_{32}\\
&+\Omega_{d}\hat{\sigma}_{54}+\text{H.c.}).
\end{split}
\end{equation}
$\hat{\sigma}_{ij} = |i\rangle\langle j|$ is the atomic operator. The detuning for each field is defined, respectively, as $\Delta_{p}=\omega_{p}-\omega_{41}, \Delta_{c}=\omega_{c}-\omega_{31}$ and $\Delta_{d}=\omega_{d}-\omega_{54}$. 
We assume that the coupling field always resonates with its transition, that is $\Delta_c = 0$, here and in the following discussions. 
The detunings of the generated Stokes and anti-Stokes fields are given respectively by $\Delta_{s}=\omega_{s}-\omega_{42}$ and $\Delta_{as}=\omega_{as}-\omega_{31}$, while
\[\omega=\Delta_{p}-\Delta_{s},\] and it is the double-photon detuning between the Stokes photon and pumping field. Due to the law of conservation of energy, one finds that $\Delta_{c} - \Delta_{as} = -\omega$.
The parameter $\delta$ is defined as 
\[\delta=\Delta_{p}+\Delta_{d},\]
and it represents the double-photon detuning between the pumping and driving field on transition $|1\rangle$ and $|5\rangle$.
The coupling strength between the atom and the cavity modes is  
$g_{\alpha} = \mu_{ij}E_{\alpha}/2\hbar$, $(\alpha \in \{s,as\})$ where $\mu_{ij}$ is the dipole moment of the corresponding transition, and $E_{\alpha}$ is the single-photon electric field, written as $E_{\alpha} = (\hbar \omega_\alpha / 2 \epsilon_0 V)^{1/2}$. $V$ is the volume of the cavity and we assume that both cavities have the same volume. The Rabi frequencies of the applied lasers are $\Omega_{\beta} = \mu_{ij}E_{\beta}/2\hbar$ with $E_\beta$ being the corresponding electric field. 

By including the coupling effect from the environment, the evolution of the system can be described by the following master equation,
\begin{equation}\label{eq.master}
\frac{\partial}{\partial t}\rho=\frac{i}{\hbar}[\rho,\hat{V}]+\mathcal{L}_{atom}(\rho)+\mathcal{L}_{s}(\rho)+\mathcal{L}_{as}(\rho),
\end{equation}
where $\mathcal{L}_{atom}(\rho)$ is the superoperators describing the spontaneous decay processes of the atom. There are six optical transitions driven by the free-space vacuum, and resultant decay terms are 
\begin{subequations}
\begin{equation}
\begin{split}
&\mathcal{L}_{atom}(\rho)=\\
&\Gamma_{54} (\hat{\sigma}_{45}\rho\hat{\sigma}_{54}-\frac{1}{2}\{\hat{\sigma}_{55},\rho\})+\Gamma_{53}(\hat{\sigma}_{35}\rho\hat{\sigma}_{53}-\frac{1}{2}\{\hat{\sigma}_{55},\rho\})\\
+&\Gamma_{42} (\hat{\sigma}_{24}\rho\hat{\sigma}_{42}-\frac{1}{2}\{\hat{\sigma}_{44},\rho\})+\Gamma_{41} (\hat{\sigma}_{14}\rho\hat{\sigma}_{41}-\frac{1}{2}\{\hat{\sigma}_{44},\rho\})\\
+&\Gamma_{32} (\hat{\sigma}_{23}\rho\hat{\sigma}_{32}-\frac{1}{2}\{\hat{\sigma}_{33},\rho\})+\Gamma_{31} (\hat{\sigma}_{13}\rho\hat{\sigma}_{31}-\frac{1}{2}\{\hat{\sigma}_{33},\rho\});
\end{split}
\end{equation}
$ \mathcal{L}_{s}(\rho)$ and $\mathcal{L}_{as}(\rho)$ represent the photon leaking processes of the Stokes and anti-Stokes photons, respectively, and they are
\begin{equation}\notag
\mathcal{L}_{s}(\rho)=\kappa_{s}(\hat{a}_{s}\rho\hat{a}^{\dagger}_{s}-\frac{1}{2}\hat{a}^{\dagger}_{s}\hat{a}_{s}\rho-\frac{1}{2}\rho\hat{a}^{\dagger}_{s}\hat{a}_{s});    
\end{equation}
\begin{equation}
\mathcal{L}_{as}(\rho)=\kappa_{as}(\hat{a}_{as}\rho\hat{a}^{\dagger}_{as}-\frac{1}{2}\hat{a}^{\dagger}_{as}\hat{a}_{as}\rho-\frac{1}{2}\rho\hat{a}^{\dagger}_{as}\hat{a}_{as}).
\end{equation}
\end{subequations}
$\Gamma_{ij}$ is the decay rate of the atomic transition $|i\rangle\leftrightarrow|j\rangle$. $\kappa_{s}$ and $\kappa_{as}$ are the leaking rates of the photons from the corresponding cavities. In the following discussion, we set $\Gamma_{54}=\Gamma_{53}=\Gamma_{1}$ and $\Gamma_{42}=\Gamma_{41}=\Gamma_{32}=\Gamma_{31}=\Gamma_{2}$, and assume that both cavities have the same leaking rate, $\kappa_{s}=\kappa_{as}=\kappa$. 

\section{Control capacity enhanced by dark State}
\begin{figure}[htbp]
\centering
\includegraphics[width= 0.6\linewidth]{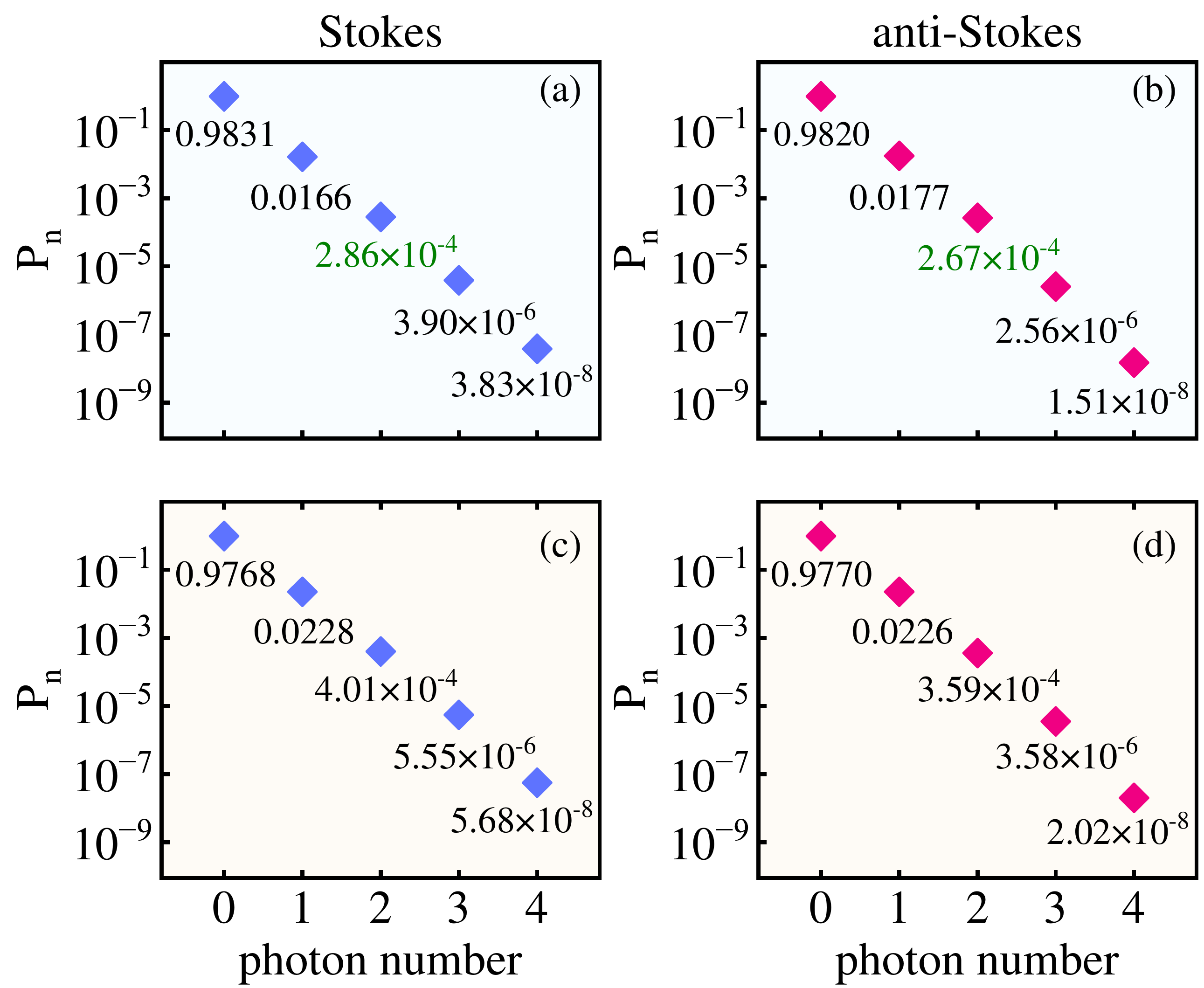}
\caption{\label{fig2} Photon number probability under puming-driving double-photon resonance $\delta = 0$ ($\Delta_{d}=-\Delta_{p}=-175\ \gamma_{31}$) for Stokes (a) and anti-Stokes (b) photons. Those under $\delta = 10\,\gamma_{31}$ ($\Delta_{p}=175\ \gamma_{31}$ and $\Delta_{d}=-165\ \gamma_{31}$) are shown in (c) and (d). The other parameters are $\Omega_{p}=10.0\, \gamma_{31},\ \Omega_{c}=1.0\,\gamma_{31},\ \Omega_{d}=2.0\,\gamma_{31}, \Delta_{c}=0,\omega=0, \kappa=0.2\,\gamma_{31},\ g=1.0\,\gamma_{31},\ \Gamma_{1}=10^{-3}\ \gamma_{31}, \Gamma_{2}=\gamma_{31}$.}
\end{figure}

Apart from the spontaneous decay processes between $|4\rangle$ and $|2\rangle$, and between $|3\rangle$ and $|1\rangle$), the system could be regarded as a combination of the subsystem $\{|1\rangle - |4\rangle - |5\rangle\}$ and $\{|2\rangle - |3\rangle\}$. As you can see that the relevant transitions are all driven by the applied fields. The double-photon resonant pumping and driving field would drive the subsystem $\{|1\rangle - |4\rangle - |5\rangle\}$ into the dark state $|\Psi_\text{dark}\rangle = (\Omega_d|1\rangle - \Omega_p |5\rangle)/ (\Omega_p^2 + \Omega_d^2)^{1/2}$
Ideally, the system would be trapped in such a state. In the following discussion we will show that it significantly changes the quantum statistical properties of the Stokes and anti-Stokes photon, and how they respond to the control provided by the driving field.

In the basis of $|j\rangle\otimes|n_{s}\rangle \otimes |n_{as}\rangle$ with $|n_{s}\rangle$ and $|n_{as}\rangle$ denoting the Fock states of the Stocks and anti-Stokes photons, and $|j\rangle$ being the $j^\text{th}$ atomic energy level, the master equation can be written as a set of equations of $\rho_{ij,m_{s},n_{s},m_{as},n_{as}}$ (see Appendix) with $\rho_{ij,m_{s},n_{s},m_{as},n_{as}} = \langle i,m_{s},m_{as}|\rho|j,n_{s},n_{as}\rangle $. In steady state ($\partial_{t}\rho_{ij,m_{s},n_{s},m_{as},n_{as}}=0$), and by truncating the Fock state at $n_{s}=n_{as}=5$, we solve the corresponding algebraic equations numerically. Based on the solutions, the photon number probability of the Stokes and anti-Stokes photon, specifically, $P_{s}(n_s)=\langle n_s|\mathrm{tr}_{as}[\rho]|n_s\rangle$ and $P_{as}(n_{as})=\langle n_{as}|\mathrm{tr}_{s}[\rho]|n_{as}\rangle$, are obtained and shown in Fig. \ref{fig2}. 
As we can see that the light field is in the vacuum state with a very high probability, and the second largest value is $P_{s/as}(1)$, which is the probability of finding one Stokes/anti-Stokes photon in the cavity. Note that for the case of double-photon resonance $\delta = 0$, $P_{as}$ is larger than $P_{s}$, see Fig. \ref{fig2}(a) and Fig. \ref{fig2}(b). Although the difference seems not large, it is clearly the opposite for $\delta = 10\,\gamma_{31}$ where $P_{s}$ is the larger.\\

\begin{figure}[htbp]
\centering
\includegraphics[width= 0.6\linewidth]{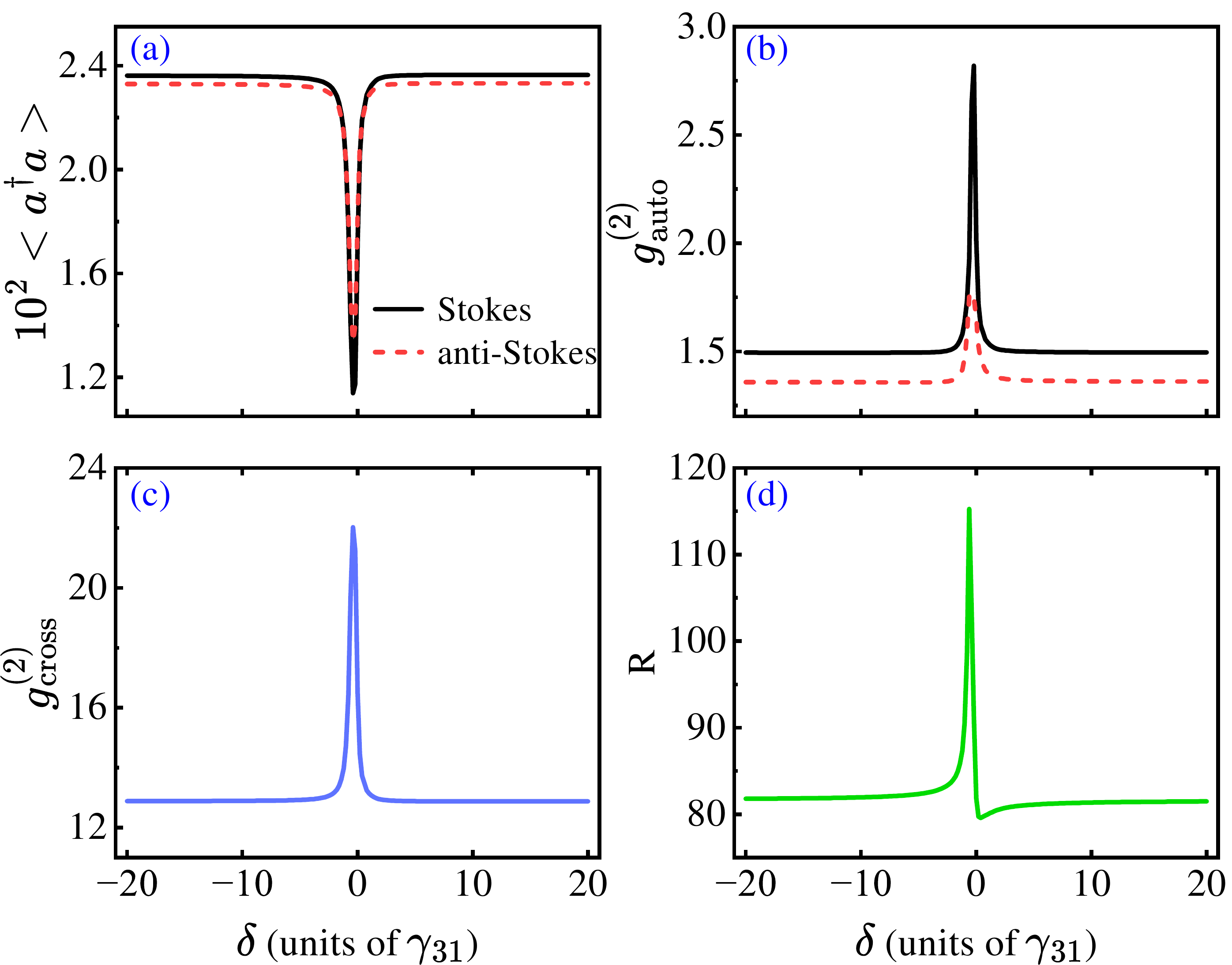}
\caption{\label{fig3}  (a) The spectral brightness, (b) degree of the auto-correlation, (c) degree of the cross-correlation and (d) ratio $R$ of the cavity-enhanced biphoton generation. Here $\Delta_{p}=175.0\, \gamma_{31}$,  $\Delta_{c}=0$, $\Omega_{p}=10.0\, \gamma_{31}$, $\Omega_{c}=1.0\, \gamma_{31}$, $\Omega_{d}=2.0\, \gamma_{31}$, $\kappa=0.2\, \gamma_{31}$, $g=1.0\, \gamma_{31}$, $\Gamma_{1}=10^{-3}\, \gamma_{31}$, $\Gamma_{2}=\gamma_{31}$ and $\omega=0$.}
\end{figure}

Such feature manifests itself in the spectral brightness ($\text{tr}[\rho a^\dagger a]$) as well. As  shown in Fig. \ref{fig3}(a) where the spectral brightness is plotted against the pumping-driving double-photon detuning $\delta$. For $\delta = 0$, the anti-Stokes brightness is larger than that of the Stokes photon, and when the pumping and driving fields are detuned from double-photon resonance either $\delta$ is positive or not, the opposite is true.
Fig. \ref{fig3}(b) shows the degree of anto-correlations, for the Stokes photon, it reads
\begin{equation}
g_{\text{auto},s}^{(2)}=\langle\hat{a}^{\dagger}_{s}\hat{a}^{\dagger}_{s}\hat{a}_{s}\hat{a}_{s}\rangle/\langle\hat{a}^{\dagger}_{s}\hat{a}_{s}\rangle\langle\hat{a}^{\dagger}_{s}\hat{a}_{s}\rangle,
\end{equation}
For the anti-Stokes photon, just replace the subscript $s$ by $as$.
Fig. \ref{fig3}(c) shows the degree of the cross-correlation, which is calculated according to
\begin{equation}
g^{(2)}_{\text{cross}}=\langle\hat{a}^{\dagger}_{as}\hat{a}^{\dagger}_{s}\hat{a}_{s}\hat{a}_{as}\rangle/\langle\hat{a}^{\dagger}_{as}\hat{a}_{as}\rangle\langle\hat{a}_{s}^{\dagger}\hat{a}_{s}\rangle.
\end{equation}
Their behavior as a function of $\delta$ is essentially the same: They all have a peak at $\delta = 0$ where the system is controlled by the dark state. Most importantly, the ratio between $g_{\text{auto}}^{(2)}$ and $g^{(2)}_{\text{cross}}$, that is
\begin{equation}
R=\frac{\left(g^{(2)}_{as,s}\right)^{2}} {g^{(2)}_{s,s}g^{(2)}_{as,as}},
\end{equation}
is commonly used to distinguish the quantum correlation from the classical: A classical source is bounded by $R\leq 1$. Fig.\ref{fig3} (d) demonstrates that the two-photon detuning $\delta$ significantly influences the nonclassical correlation strength $R$ between the Stokes and anti-Stokes photons. From the overall trend, when the system is far from the dark-state condition, the $R$ value remains relatively stable, staying around 80. However, when the system is close to entering the dark state between $|1\rangle$ and $|5\rangle$ ($\delta\approx-0.6\ \gamma_{31}$), the $R$ value shows a distinct increase, reaching a maximum of approximately 115. Nevertheless, over the entire detuning range, the $R$ value is always far greater than 1, indicating that an extremely strong nonclassical correlation exists between the Stokes and anti-Stokes photon pairs in this system.
\begin{figure}[htbp]
\centering
\includegraphics[width= 0.6\linewidth]{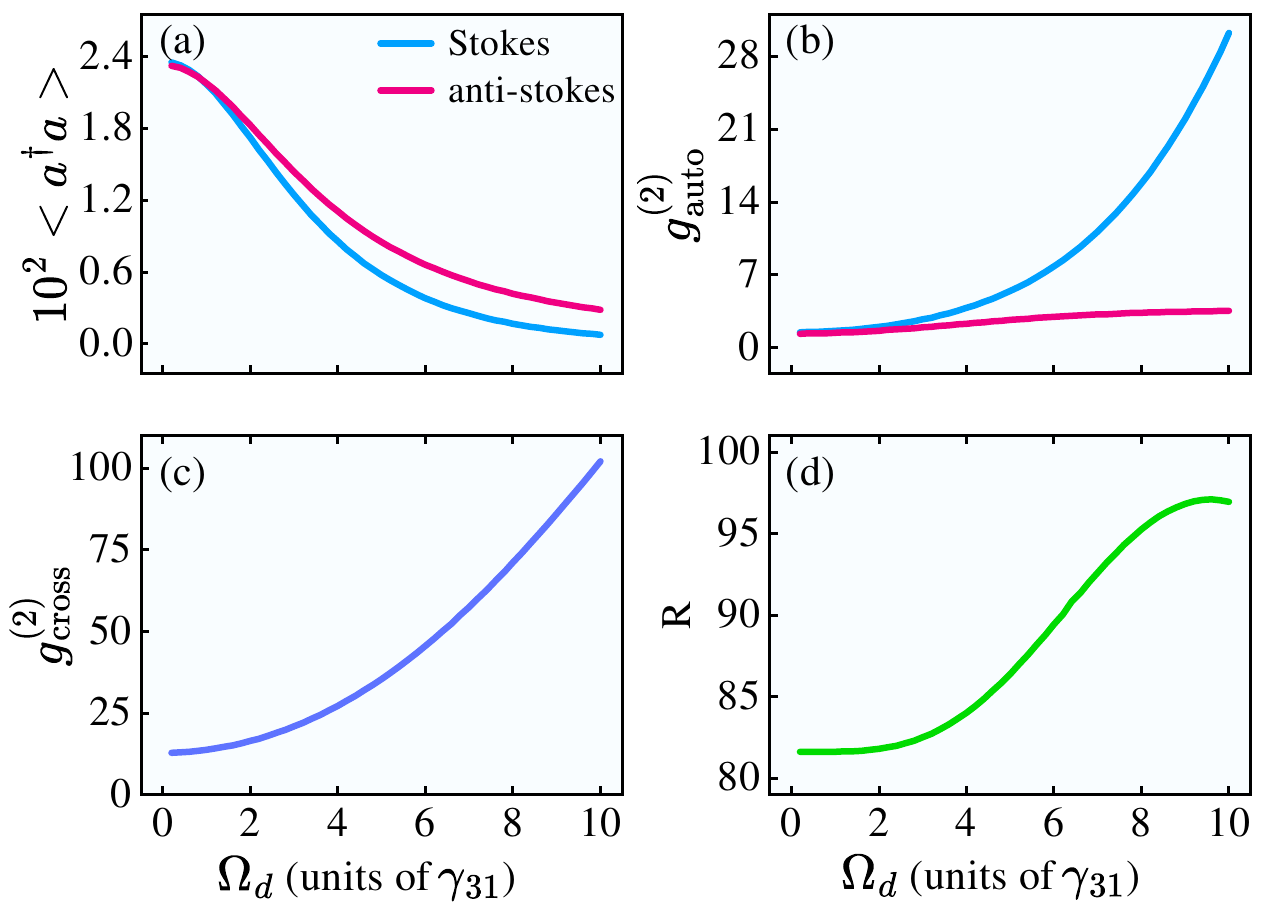}
\caption{\label{fig4}
The spectral brightness (a), degree of the auto-correlation (b), degree of the cross-correlation (c) and the ratio $R$ (d) under different driving Rabi frequencies. Pumping-driving double-photon detuning $\delta = 0$.
The other parameters are $\Delta_{c}=0$, $\omega=0$, $\Omega_{p}=10.0\,\gamma_{31}$, $\Omega_{c}=1.0\ \gamma_{31}$, $\kappa=0.2\, \gamma_{31}$, $g=1.0\,\gamma_{31}$, $\Gamma_{1}=10^{-3}\ \gamma_{31}$, $\Gamma_{2}=\gamma_{31}.$}
\end{figure}

\section{Coherent control on the biphoton generation}

In general, each of the three applied fields can influence the generation of the biphotons. The pumping field drives the transition of $|1\rangle \leftrightarrow |4\rangle$ where level $|1\rangle$ is the most populated. Increasing it would increase the generation rate, however it enhances the Raman enhancement as well. In the following discussions, we do not treat $\Omega_p$ or $\Delta_p$ as a control knob, but only focus on the influences from the driving and coupling field. Considering that the strength of the interaction between the atom and the cavity is a unique parameter of our system, we also show in this section how the generation and correlation changes with different value of $g$.

\subsection{Control via driving field}

Recall that $\delta = \Delta_p + \Delta_d$, for a fixed pumping detuning $\Delta_p$, changing the frequency of the driving field, can change the value of $\delta$. As suggested by Fig. \ref{fig3}, for large $|\delta|$, it does not alter the statistical properties much of the generated photons if $\delta$ is changed. However, when  $\delta$ is close to 0, and the dark state is built between the ground and Rydberg state, it  certainly enhances the sensitivity of the system on the strength of the driving field. 

Fig. \ref{fig4} shows the numerical results for $\delta = 0$ with $\Delta_p = 175\,\gamma_{31}$. That corresponds to the case with $\Delta_d = -175\,\gamma_{31}$. As we can see that when the strength of the driving field is increased, both Stokes and anti-Stokes brightness are reduced. In detail, the brightness of the Stokes photons drops more quickly than that of anti-Stokes. Fig. \ref{fig4}(b) shows that the auto-correlation of the Stokes field is more sensitive to $\Omega_d$ than that of the anti-Stokes field. 
$g_{\text{auto},s}$ increases with the Rabi frequency of the driving field, while that of the anti-Stokes remains constant.
The degree of the cross-correlation is increased. That leads to an increasing ratio $R$, suggesting that the single-atom system behaves more like a quantum-correlated light source if a stronger driving field is applied. 

\begin{figure}[htbp]
\centering
\includegraphics[width= 0.65\linewidth]{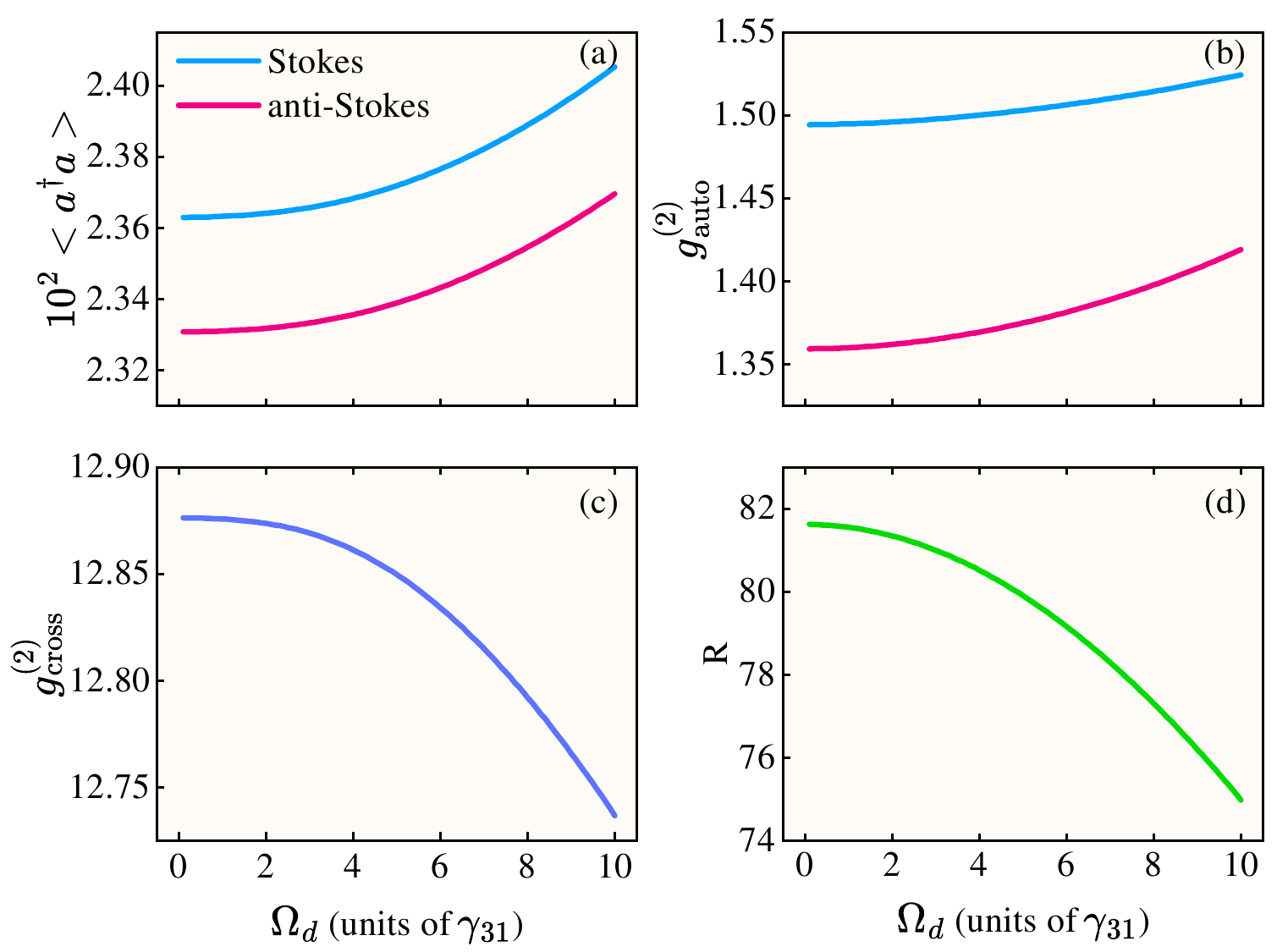}
\caption{\label{fig5}The spectral brightness (a), degree of the auto-correlation (b), degree of the cross-correlation (c) and the ratio $R$ (d) under different driving Rabi frequencies. Pumping-driving double-photon detuning $\delta = 10 \, \gamma_{31} (\Delta_{p}=175\ \gamma_{31},\ \Delta_{d}=-165\ \gamma_{31})$. The other parameters are $\Delta_{c}=0$, $\omega=0$, $\Omega_{p}=10.0\,\gamma_{31}$, $\Omega_{c}=1.0\ \gamma_{31}$, $\kappa=0.2\, \gamma_{31}$, $g=1.0\,\gamma_{31}$, $\Gamma_{1}=10^{-3}\ \gamma_{31}$, $\Gamma_{2}=\gamma_{31}.$}
\end{figure}

 Under the pump-driving double-photon off-resonant condition (\(\delta = 10\ \gamma_{31}\)), Fig. \ref{fig5} reveals a set of quantum statistical properties distinctly different from the double-photon resonance case shown in Fig. \ref{fig4}. As illustrated in Fig. \ref{fig5} (a), the spectral brightness of both Stokes and anti-Stokes photons increases slowly and synchronously with the driving Rabi frequency \(\Omega_d\), with the Stokes brightness consistently exceeding that of the anti-Stokes counterpart. In Fig. \ref{fig5} (b), the auto-correlation \(g^{(2)}_{\text{auto}}\) remains greater than unity for all \(\Omega_d\), indicating persistent photon bunching, and the auto-correlation of the Stokes photons is always higher than that of the anti-Stokes photons. Regarding cross-correlation, unlike the substantial increase observed in Fig. 4 as \(\Omega_d\) increases, Fig. \ref{fig5} (c) shows a slight decline in cross-correlation under the off-resonant condition, suggesting a certain degree of robustness. Finally, Fig. \ref{fig5} (d) demonstrates that although the nonclassical correlation metric $R$ decreases with increasing $\Omega_d$, its value consistently exceeds 75, indicating a remarkably strong nonclassical correlation.

\subsection{Control via Coupling fields}

\begin{figure}[htbp]
\centering
\includegraphics[width= 0.6\linewidth]{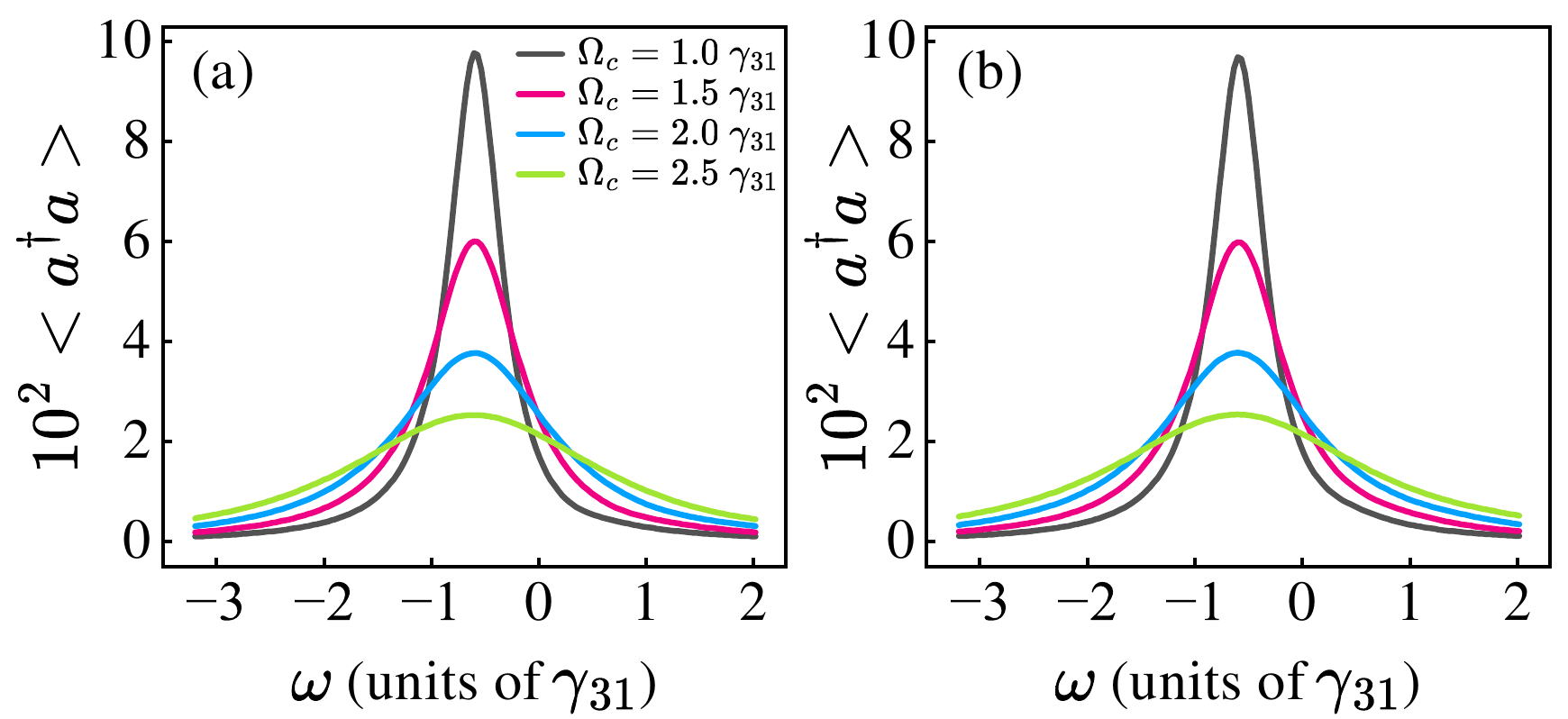}
\caption{\label{fig6} Influence of the coupling field on (a) the Stokes spectral brightness and (b) anti-Stokes spectral brightness. Here $\delta=0$, $\Delta_{c}=0$,  $\Omega_{p}=10.0\,\gamma_{31}$, $\Omega_{d}=2.0\,\gamma_{31}$, $\kappa=0.2 \,  \gamma_{31}$, $g=1.0\,\gamma_{31}$, $\Gamma_{1}=10^{-3}\,\gamma_{31}$, $\Gamma_{2}=\gamma_{31}$.}
\end{figure}

One important advantage of the double-$\Lambda$ scheme for the biphoton generation is the transparency of the anti-Stokes field due to the resonant coupling field. 
The width of the transparent window is determined by the strength of the coupling field \cite{RevModPhys.77.633}. Fig. \ref{fig6}(a) shows the brightness of the Stokes photon as a function of the pumping-Stokes double-photon detuning $\omega$. 
Lines in Fig. \ref{fig6}(a) reflect how the bandwidth of the generated photon changes with respect to the increasing coupling intensity. Note that  
the coupling Rabi frequency is proportional to the square root of the intensity of the coupling field. 

As \( \Omega_c \) increases, the EIT window broadens, and the full width at half maximum (FWHM) of the photon pair peak increases accordingly. Considering that a narrower linewidth helps enhance the nonlinear interaction between atoms and the light field, thereby improving the generation efficiency of photon pairs, our choice of \(\Omega_c = 1.0\ \gamma_{31}\) in the preceding discussion is reasonable. This parameter ensures a relatively high peak brightness while maintaining a relatively narrow linewidth. Fig. \ref{fig6}(b) shows the brightness of the anti-Stokes photon. As we can see that it is quite similar to the Stokes photon, which is actually the result of the energy conservation. 

\subsection{Cavity parameter}

\begin{figure}[htbp]
\centering
\includegraphics[width= 0.6\linewidth]{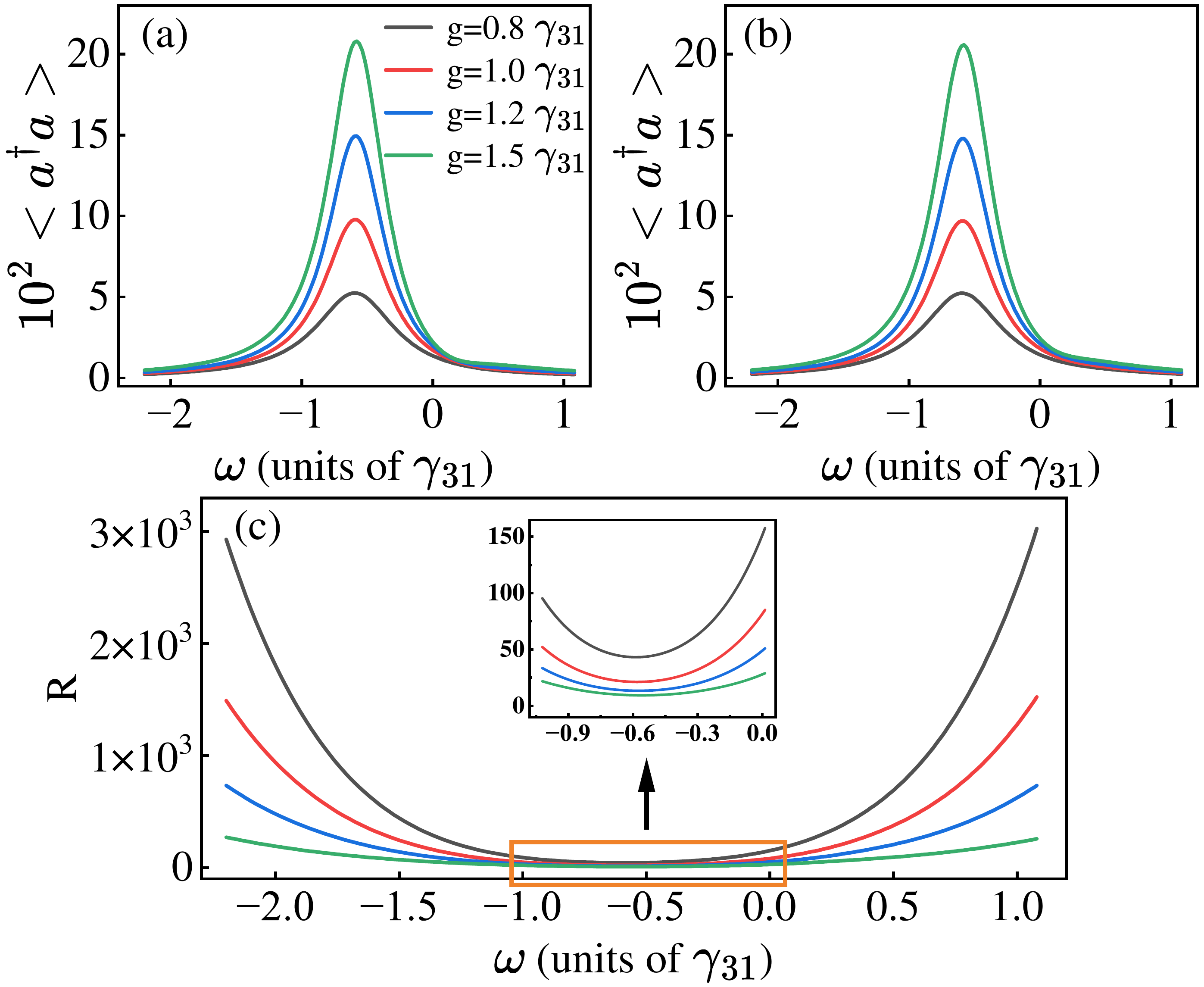}
\caption{\label{fig7} (a) Stokes spectral brightness, (b) anti-Stokes spectral brightness, (c) ratio of the cavity-enhanced biphoton generation. Here $\delta=0,\ \Delta_{c}=0,\ \ \Omega_{p}=10.0\ \gamma_{31},\ \Omega_{d}=2.0,\ \Omega_{c}=1.0\ \gamma_{31},\ \kappa=0.2\ \gamma_{31},\ \Gamma_{1}=10^{-3}\ \gamma_{31},\ \Gamma_{2}=\ \gamma_{31}.$}
\end{figure}

The coupling strength between the atom and the cavity mode is represented by $g$ and depends on the effective volume of the cavity. In our model we assume that both cavities have the same volume $V$. This might not be a parameter that can be easily changed for a given experimental setup. However, as a representative parameter in the cavity-mode-enhanced SFWM process, the effect of $g$ is certainly worth investigating, and the results show that it changes the quantum statistics of the biphotons effectively. In Fig. \ref{fig7}, we show the spectral brightness of the Stokes and anti-Stokes photon in Panel (a) and (b), respectively. The ratio $R$ for different values of $g$ is shown in Panel (c). As we can see that for a fixed coupling Rabi frequency, the linewidth of the biphoton remains almost unchanged when g increases from $0.8 \, \gamma_{31}$ to $1.5\,\gamma_{31}$. However the maximal value of it increases almost linearly with respect to $g$. When the brightness is reduced, for a relatively small value of $g$, the ratio $R$ increases. This principle applies not only to the comparison of the lines in Fig. \ref{fig7} for different $g$, but also to a single line with certain g and varying $\omega$. Smaller volume is favored for higher brightness of the biphotons but is not helpful if the nonclassical statistics is a necessary requirement.

\section{Conclusions}

In this paper, we have studied the coherent control of biphoton generation in a cavity-enhanced Rydberg-atom system based on spontaneous four-wave mixing. When the pumping and driving fields satisfy the two-photon resonance condition ($\delta = 0$), the system is trapped in a dark state between the ground state and the Rydberg state. In this regime, the Rabi frequency of the driving field can significantly modify the spectral brightness, auto-correlation, and cross-correlation of both Stokes and anti-Stokes photons, demonstrating efficient coherent manipulation. On the other hand, when the two-photon detuning is large, the control effect of the driving field becomes much weaker.
The coupling field plays a crucial role through the electromagnetically induced transparency effect. Increasing the coupling Rabi frequency broadens the EIT window and thereby increases the linewidth of the generated biphotons. A weaker coupling field is favorable for achieving a narrower linewidth, which is beneficial for nonlinear interactions.
The atom-cavity coupling strength $g$, which depends on the cavity volume, directly influences the brightness of the biphotons: a larger 
$g$ (smaller cavity volume) gives higher peak brightness, while the linewidth remains almost unchanged. However, a smaller 
$g$ (larger volume) tends to enhance the nonclassical correlation ratio 
$R$, indicating a trade-off between brightness and quantum statistics.
Finally, under all relevant parameter conditions examined, the Stokes and anti-Stokes photons exhibit strong quantum correlations, with the ratio 
$R$ always significantly exceeding unity.

\appendix

\section{THE EQUATIONS OF THE ELEMENT OF THE DENSITY MATRIX}\label{AppA}

We define the density matrix element  $\langle i,m_{s},m_{as}|\rho|j,n_{s},n_{as}\rangle$ as $\rho_{ij,m_{s},n_{s},m_{as},n_{as}}$, where $i,j$ denote the atoms levels, and $|m_{s}\rangle$ and $|m_{as}\rangle$ are the Fock basis of the photon pair. Based on the interaction Hamiltonian and the master equation, a differential equation with $\rho_{ij,m_{s},n_{s},m_{as},n_{as}}$ as the unknowns can be derived. In these equations, $\gamma_{ij}$ denotes the dephasing rate due to the interaction with the environment ($\mathcal{L}_{atom}$), while $
\mathcal{D}(\rho_{ij,m_{s},n_{s},m_{as},n_{as}})$ represents the dephasing processes arising from photon leaking in the two cavities ($\mathcal{L}_{s},\mathcal{L}_{as}$). Due to the assumption of an ideal laser, additional decoherence terms such as those arising from the finite linewidths of the pump and coupling fields have been neglected.

Based on the master equation (2), the elements of the density matrix obey the following equations:
\begin{equation}
\begin{split}
\frac{d}{dt}\rho_{11;m_{s},n_{s};m_{as},n_{as}}&=\Gamma_{41}\rho_{44;m_{s},n_{s};m_{as},n_{as}}+\Gamma_{31}\rho_{33;m_{s},n_{s};m_{as},n_{as}}\\
&-i(g_{as}\sqrt{n_{as}}\rho_{13;m_{s},n_{s};m_{as},n_{as}-1}+\Omega_{p}\rho_{14;m_{s},n_{s};m_{as},n_{as}}\\
&-g_{as}\sqrt{m_{as}}\rho_{31;m_{s},n_{s};m_{as}-1,n_{as}}-\Omega_{p}^{\ast}\rho_{41;m_{s},n_{s};m_{as},n_{as}})
\end{split}
\end{equation}
\begin{equation}
\begin{split}
\frac{d}{dt}\rho_{22;m_{s},n_{s};m_{as},n_{as}}&=\Gamma_{42}\rho_{44;m_{s},n_{s};m_{as},n_{as}}+\Gamma_{32}\rho_{33;m_{s},n_{s};m_{as},n_{as}}\\
&-i(g_{s}\sqrt{n_{s}}\rho_{24;m_{s},n_{s}-1;m_{as},n_{as}}+\Omega_{c}\rho_{23;m_{s},n_{s};m_{as},n_{as}}\\
&-g_{s}\sqrt{m_{s}}\rho_{42;m_{s}-1,n_{s};m_{as},n_{as}}-\Omega_{c}^{\ast}\rho_{32;m_{s},n_{s};m_{as},n_{as}})
\end{split}
\end{equation}
\begin{equation}
\begin{split}
\frac{d}{dt}\rho_{33;m_{s},n_{s};m_{as},n_{as}}&=\Gamma_{53}\rho_{55;m_{s},n_{s};m_{as},n_{as}}-(\Gamma_{32}+\Gamma_{31})\rho_{33;m_{s},n_{s};m_{as},n_{as}}\\
&-i(g_{as}\sqrt{n_{as}+1}\rho_{31;m_{s},n_{s};m_{as},n_{as}+1}+\Omega_{c}^{\ast}\rho_{32;m_{s},n_{s};m_{as},n_{as}}\\
&-g_{as}\sqrt{m_{as}+1}\rho_{13;m_{s},n_{s};m_{as}+1,n_{as}}-\Omega_{c}\rho_{23;m_{s},n_{s};m_{as},n_{as}})
\end{split}
\end{equation}
\begin{equation}
\begin{split}
\frac{d}{dt}\rho_{44;m_{s},n_{s};m_{as},n_{as}}&=\Gamma_{54}\rho_{55;m_{s},n_{s};m_{as},n_{as}}-(\Gamma_{42}+\Gamma_{41})\rho_{44;m_{s},n_{s};m_{as},n_{as}}\\
&-i(g_{s}\sqrt{n_{s}+1}\rho_{42;m_{s},n_{s}+1;m_{as},n_{as}}+\Omega_{d}\rho_{45;m_{s},n_{s};m_{as},n_{as}}\\
&+\Omega_{p}^{\ast}\rho_{41;m_{s},n_{s};m_{as},n_{as}}-g_{s}\sqrt{m_{as}+1}\rho_{24;m_{s}+1,n_{s};m_{as},n_{as}}\\
&-\Omega_{d}^{\ast}\rho_{54;m_{s},n_{s};m_{as},n_{as}}-\Omega_{p}\rho_{14;m_{s},n_{s};m_{as},n_{as}})
\end{split}
\end{equation}
\begin{equation}
\begin{split}
\frac{d}{dt}\rho_{55;m_{s},n_{s};m_{as},n_{as}}&=-(\Gamma_{54}+\Gamma_{53})\rho_{55;m_{s},n_{s};m_{as},n_{as}}\\
&-i(\Omega_{d}^{\ast}\rho_{54;m_{s},n_{s};m_{as},n_{as}}-\Omega_{d}\rho_{45;m_{s},n_{s};m_{as},n_{as}})
\end{split}
\end{equation}
\begin{equation}
\begin{split}
\frac{d}{dt}\rho_{54;m_{s},n_{s};m_{as},n_{as}}&=-(\gamma_{54}-i\Delta_{d})\rho_{54;m_{s},n_{s};m_{as},n_{as}}\\
&-i(g_{s}\sqrt{n_{s}+1}\rho_{52;m_{s},n_{s}+1;m_{as},n_{as}}+\Omega_{d}\rho_{55;m_{s},n_{s};m_{as},n_{as}}\\
&+\Omega_{p}^{\ast}\rho_{{51};m_{s},n_{s};m_{as},n_{as}}-\Omega_{d}\rho_{44;m_{s},n_{s};m_{as},n_{as}})
\end{split}
\end{equation}
\begin{equation}
\begin{split}
\frac{d}{dt}\rho_{53;m_{s},n_{s};m_{as},n_{as}}&=-[\gamma_{53}+i(\omega-\delta)]\rho_{53;m_{s},n_{s};m_{as},n_{as}}\\
&-i(g_{as}\sqrt{n_{as}+1}\rho_{51;m_{s},n_{s};m_{as},n_{as}+1}+\Omega_{c}^{\ast}\rho_{52;m_{s},n_{s};m_{as},n_{as}}\\
&-\Omega_{d}\rho_{43;m_{s},n_{s};m_{as},n_{as}})
\end{split}
\end{equation}
\begin{equation}
\begin{split}
\frac{d}{dt}\rho_{52;m_{s},n_{s};m_{as},n_{as}}&=-[\gamma_{52}+i(\omega-\delta)]\rho_{52;m_{s},n_{s};m_{as},n_{as}}\\
&-i(g_{s}\sqrt{n_{s}}\rho_{54;m_{s},n_{s}-1;m_{as},n_{as}}+\Omega_{c}\rho_{53;m_{s},n_{s};m_{as},n_{as}}\\
&-\Omega_{d}\rho_{42;m_{s},n_{s};m_{as},n_{as}})
\end{split}
\end{equation}
\begin{equation}
\begin{split}
\frac{d}{dt}\rho_{51;m_{s},n_{s};m_{as},n_{as}}&=-(\gamma_{51}-i\delta)\rho_{51;m_{s},n_{s};m_{as},n_{as}}\\
&-i(g_{as}\sqrt{n_{as}}\rho_{53;m_{s},n_{s};m_{as},n_{as}-1}+\Omega_{p}\rho_{54;m_{s},n_{s};m_{as},n_{as}}\\
&-\Omega_{d}\rho_{41;m_{s},n_{s};m_{as},n_{as}})
\end{split}
\end{equation}
\begin{equation}
\begin{split}
\frac{d}{dt}\rho_{43;m_{s},n_{s};m_{as},n_{as}}&=-[\gamma_{43}+i(\omega-\Delta_{p})]\rho_{43;m_{s},n_{s};m_{as},n_{as}}\\
&-i(g_{as}\sqrt{n_{as}+1}\rho_{41;m_{s},n_{s};m_{as},n_{as}+1}+\Omega_{c}^{\ast}\rho_{42;m_{s},n_{s};m_{as},n_{as}}\\
&-g_{s}\sqrt{m_{s}+1}\rho_{23;m_{s}+1,n_{s};m_{as},n_{as}}-\Omega_{d}^{\ast}\rho_{53;m_{s},n_{s};m_{as},n_{as}}\\
&-\Omega_{p}\rho_{13;m_{s},n_{s};m_{as},n_{as}})
\end{split}
\end{equation}
\begin{equation}
\begin{split}
\frac{d}{dt}\rho_{42;m_{s},n_{s};m_{as},n_{as}}&=-[\gamma_{42}+i(\omega-\Delta_{p})]\rho_{42;m_{s},n_{s};m_{as},n_{as}}\\
&-i(g_{s}\sqrt{n_{s}}\rho_{44;m_{s},n_{s}-1;m_{as},n_{as}}+\Omega_{c}\rho_{43;m_{s},n_{s};m_{as},n_{as}}\\
&-g_{s}\sqrt{m_{s}+1}\rho_{22;m_{s}+1,n_{s};m_{as},n_{as}}-\Omega_{p}\rho_{12;m_{s},n_{s};m_{as},n_{as}}\\
&-\Omega_{d}^{\ast}\rho_{52;m_{s},n_{s};m_{as},n_{as}})
\end{split}
\end{equation}
\begin{equation}
\begin{split}
\frac{d}{dt}\rho_{41;m_{s},n_{s};m_{as},n_{as}}&=-(\gamma_{41}-i\Delta_{p})\rho_{41;m_{s},n_{s};m_{as},n_{as}}\\
&-i(g_{as}\sqrt{n_{as}}\rho_{43;m_{s},n_{s};m_{as},n_{as}-1}+\Omega_{p}\rho_{44;m_{s},n_{s};m_{as},n_{as}}\\
&-g_{s}\sqrt{m_{s}+1}\rho_{21;m_{s}+1,n_{s};m_{as},n_{as}}-\Omega_{p}\rho_{11;m_{s},n_{s};m_{as},n_{as}}\\
&-\Omega_{d}^{\ast}\rho_{51;m_{s},n_{s};m_{as},n_{as}})
\end{split}
\end{equation}
\begin{equation}
\begin{split}
\frac{d}{dt}\rho_{32;m_{s},n_{s};m_{as},n_{as}}&=-\gamma_{32}\rho_{32;m_{s},n_{s};m_{as},n_{as}}\\
&-i(g_{s}\sqrt{n_{s}}\rho_{34;m_{s},n_{s}-1;m_{as},n_{as}}+\Omega_{c}\rho_{33;m_{s},n_{s};m_{as},n_{as}}\\
&-g_{as}\sqrt{m_{as}+1}\rho_{12;m_{s},n_{s};m_{as}+1,n_{as}}-\Omega_{c}\rho_{22;m_{s},n_{s};m_{as},n_{as}})
\end{split}
\end{equation}
\begin{equation}
\begin{split}
\frac{d}{dt}\rho_{31;m_{s},n_{s};m_{as},n_{as}}&=-(\gamma_{31}-i\omega)\rho_{31;m_{s},n_{s};m_{as},n_{as}}\\
&-i(g_{as}\sqrt{n_{as}}\rho_{33;m_{s},n_{s};m_{as},n_{as}-1}+\Omega_{p}\rho_{34;m_{s},n_{s};m_{as},n_{as}}\\
&-g_{as}\sqrt{m_{as}+1}\rho_{11;m_{s},n_{s};m_{as}+1,n_{as}}-\Omega_{c}\rho_{21;m_{s},n_{s};m_{as},n_{as}})
\end{split}
\end{equation}
\begin{equation}
\begin{split}
\frac{d}{dt}\rho_{21;m_{s},n_{s};m_{as},n_{as}}&=i\omega\rho_{21;m_{s},n_{s};m_{as},n_{as}}\\
&-i(g_{as}\sqrt{n_{as}}\rho_{23;m_{s},n_{s};m_{as},n_{as}-1}+\Omega_{p}\rho_{24;m_{s},n_{s};m_{as},n_{as}}\\
&-g_{s}\sqrt{m_{s}-1}\rho_{41;m_{s}-1,n_{s};m_{as},n_{as}}-\Omega_{c}^{\ast}\rho_{31;m_{s},n_{s};m_{as},n_{as}})
\end{split}
\end{equation}
In each equation, there is a hidden term resulted from the $\mathcal{L}_{cav}(\rho)$ which is
\begin{equation}
\begin{split}
\mathscr{D}(\rho_{\alpha\beta;m_{s}n_{s};m_{as},n_{as}})=&-\frac{\kappa_{s}}{2}(m_{s}+n_{s})\rho_{\alpha\beta;m_{s}n_{s};m_{as},n_{as}}\\
&+\kappa_{s}\sqrt{{(m_{s}+1})(n_{s}+1)}\rho_{\alpha\beta;m_{s}+1,n_{s}+1;m_{as},n_{as}}\\
&
-\frac{\kappa_{as}}{2}(m_{as}+n_{as})\rho_{\alpha\beta;m_{s}n_{s};m_{as},n_{as}}\\
&+\kappa_{as}\sqrt{{(m_{as}+1})(n_{as}+1)}\rho_{\alpha\beta;m_{s}n_{s};m_{as}+1,n_{as}+1}
\end{split}
\end{equation}
The atomic decoherence rates $\gamma_{ij}$ are
$\gamma_{54}=\frac{1}{2}(\Gamma_{54}+\Gamma_{53}+\Gamma_{42}+\Gamma_{41})$,
$\gamma_{53}=\frac{1}{2}(\Gamma_{54}+\Gamma_{53}+\Gamma_{32}+\Gamma_{31})$,
$\gamma_{52}=\gamma_{51}=\frac{1}{2}(\Gamma_{54}+\Gamma_{53})$,
$\gamma_{43}=\frac{1}{2}(\Gamma_{42}+\Gamma_{41}+\Gamma_{32}+\Gamma_{31})$,
$\gamma_{42}=\gamma_{41}=\frac{1}{2}(\Gamma_{42}+\Gamma_{41})$,
$\gamma_{32}=\gamma_{31}=\frac{1}{2}(\Gamma_{32}+\Gamma_{31})$.

\begin{backmatter}
\bmsection{Funding}
National Natural Science
Foundation of China (No. 62375047)

\bmsection{Acknowledgment}
The work is supported by the National Natural Science
Foundation of China (No. 62375047).

\bmsection{Disclosures}
The authors declare no conflicts of interest.

\bmsection{Data availability} Data underlying the results presented in this paper are not publicly available at this time but may
be obtained from the authors upon reasonable request.

\end{backmatter}


\bibliography{sample}






\end{document}